\documentclass[final,3p,times,sort&compress]{elsarticle}

\usepackage{amsmath,amssymb,amsfonts,graphicx,slashed,citesort,dsfont,ulem}
\usepackage{array,multirow,booktabs}
\usepackage[usenames]{color}
\usepackage{enumitem}
\allowdisplaybreaks[1]
\newcommand{\nn}{\nonumber}
\newcommand{\eq}{\begin{eqnarray}}
\newcommand{\en}{\end{eqnarray}}
\newcommand{\Gr}{G_{\rm r}}

\begin{document}

\begin{frontmatter}
\title{Recoil corrections in antikaon-deuteron scattering}

\author[Bonn]{Maxim~Mai}
\author[Bochum,Moscow]{Vadim~Baru}
\author[Bochum]{Evgeny~Epelbaum}
\author[Bonn]{Akaki~Rusetsky}

\address[Bonn]{Helmholtz--Institut f\"ur Strahlen- und Kernphysik (Theorie) and Bethe Center for Theoretical Physics, Universit\"at Bonn, D-53115 Bonn, Germany}
\address[Bochum]{Institut f\"ur theoretische Physik II, Fakult\"at f\"ur Physik und Astronomie, Ruhr-Universit\"at Bochum, 44780 Bochum, Germany}
\address[Moscow]{Institute for Theoretical and Experimental Physics, 117218, B. Cheremushkinskaya 25, Moscow, Russia}

\begin{abstract} 
The recoil retardation effect in the $K^-d$ scattering length is studied. Using the
non-relativistic effective field theory approach, it is demonstrated that
a systematic {\it perturbative} expansion of the recoil corrections
in the parameter $\xi=M_K/m_N$
is possible in spite of the fact that $K^-d$ scattering at low energies is
inherently non-perturbative due to the large values of the $\bar KN$ scattering
lengths.  The first order correction to the $K^-d$ scattering length due to single insertion of the retardation term  in the multiple-scattering series
is  calculated.  The recoil effect  turns out to be reasonably small even
 at the physical value of $M_K/m_N\simeq 0.5$.

\end{abstract}

\end{frontmatter}

\section{Introduction}

Antikaon-nucleon scattering is an excellent testing ground for
 understanding the $SU(3)$ QCD dynamics at low energies in the one-baryon
 sector. Starting from the seminal paper~\cite{Siegel}, it is often described
 within the so-called unitarized Chiral Perturbation Theory (ChPT),
which uses the chiral potential calculated at a certain order (see, e.g.,
 Refs.~\cite{OsetRamos,MO,Nissler,Oller,BMN,Jido,Lutz,Ikeda:2012au,Mai:2012dt}).
Different versions of the unitarized ChPT are available in literature.
However, a common feature of all formulations is a relatively large number of free
 parameters, which are fixed from the fit to the experimental data. 
An important part of the 
input is coming from the S-wave $\bar KN$ scattering lengths, 
which ``nail down'' the amplitudes at the $\bar KN$ 
threshold and thus impose stringent
constraints both on the scattering in the $\bar KN$ channel as well as
the sub-threshold behavior of the amplitudes. The latter plays an important role
in the study of the interaction of the $K^-$ with nuclear medium (see, e.g.,
Ref.~\cite{Weise:2005ss}).  

The experiments with kaonic atoms have been carried out in order to extract
the precise values of the S-wave $\bar KN$ scattering lengths -- a goal that
could be hardly achieved by using different experimental techniques.
Recently, the
 energy shift and width of kaonic hydrogen were measured very accurately in 
the SIDDHARTA experiment at  {DA$\Phi$NE~\cite{Bazzi:2011zj}} (for the
earlier attempts, see, e.g., Refs.~\cite{KEK,Zmeskal}). These two 
quantities can be related to the $K^-p$ scattering lengths via the 
so-called modified Deser-type formula, see 
Refs.~\cite{Deser:1954vq,Meissner:2004jr} (a general discussion of
the procedure of 
extracting the scattering lengths from the hadronic atom observables
can be found, e.g., in a review article~\cite{physrep}).
The same experimental collaboration has made an attempt to measure the
energy and the width of the ground state of the kaonic deuterium as well.
However, due to a small signal-to-background ratio, no clear signal
from the kaonic deuterium was detected~\cite{Ishiwatari}. Only an upper limit
on the yield of the kaonic deuterium $K_\alpha$ line could be
determined which is important for the evaluation of the new experiments proposed at 
LNF~\cite{LNF} and J-PARC~\cite{JPARK}. In addition, 
the kaonic $^3He$ and $^4He$ 
atoms have also been studied within the SIDDHARTA experiment.

What makes the experiments with   kaonic deuterium extremely important
is the fact that the S-wave $\bar KN$ scattering lengths  are complex-valued, in difference, e.g.,  to the $\pi N$ scattering
lengths, which are real quantities.     
The $\pi N$ scattering lengths  can be, in principle, determined directly on the basis of  $\pi^- H$  data alone~\cite{Gottawidth} while  the level shift of 
the pionic  deuterium~\cite{Strauch:2010vu} is used as a complementary information  to improve  the accuracy in the  extraction of the    $\pi N$ scattering lengths, see 
Ref.~\cite{JOB} for the results  of the latest  combined  analysis  of pionic atoms. 
Meanwhile, extracting two complex scattering lengths $a_0,a_1$ 
corresponding to the total isospin $I=0,1$ in the $\bar K N$  system implies the determination of four
real quantities and thus requires   measurements of four independent 
observables. Two observables are provided by the kaonic hydrogen, and the
remaining two can come from, e.g., the kaonic deuterium. The  
problem is, however, that the kaonic deuterium measurement would yield the
$\bar Kd$ scattering length, and there is still a long 
way to go from this quantity to the $\bar KN$ scattering lengths. Thus, unless one is able to relate the
$\bar Kd$ and $\bar KN$ scattering lengths with each other with a controlled accuracy 
using a consistent theoretical framework, the main goal of the 
kaonic deuterium experiment can not be achieved.

In the past decades, the calculations of the 
low-energy $\bar Kd$ scattering observables within the non-relativistic
three-body Faddeev framework have reached an unprecedented accuracy and 
sophistication (see here an incomplete list of papers on this 
subject:~\cite{Hetherington,Torres:1986mr,toker,mizutani,Shevchenko:2011ce,Shevchenko:2012np,Shevchenko:2014uva}). 
However, despite all efforts, these calculations do not address the core 
issue, which consists in the feasibility of the extraction of the $\bar KN$ 
scattering lengths from data. To this end, one needs an explicit expression  
of the $\bar Kd$ scattering length in terms of the $\bar KN$ scattering
lengths of the type of  Brueckner  formula~\cite{Chand}, see also 
Ref.~\cite{Kamalov}. Based on this expression, in 
Refs.~\cite{Meissner:2006gx,Doring:2011xc} the extraction of the $\bar KN$ 
scattering lengths from the combined data on the kaonic hydrogen and 
kaonic deuterium has been analyzed. It should be however stressed that the
expression derived in Refs.~\cite{Chand,Kamalov,Meissner:2006gx} is only
an approximation and assumes that nucleons are infinitely heavy (static).
There exists no {\it a priori} reason to believe that this is a
good approximation, especially in  view of the fact that, in the real world,
the mass ratio $M_K/m_N\simeq 0.5$ is not small. On the other hand,
{\it numerical} estimates, carried out within the Faddeev approach, indicate that
this approximation might be not so bad as it seems at the first glance
(see Ref.~\cite{Gal}). One therefore may conclude that,
 in order to have a {\it reliable} estimate
of the uncertainty of the method, it is necessary to have a systematic
framework for calculating corrections to the static limit. This issue has
been first addressed in Ref.~\cite{Baru:2009tx}, where the recoil corrections
in the double scattering process have been studied within the non-relativistic
field-theoretical approach. This paper represents a continuation and extension 
of the work  started in Ref.~\cite{Baru:2009tx}.

Below we list some fundamental questions that the  systematic theory of kaonic deuterium should be able to answer in the future:

\begin{itemize}

\item[i)]
$\bar Kd$ scattering at low energies is inherently non-perturbative
due to the large values of the S-wave $\bar KN$ scattering lengths, so a  
re-summation of the multiple scattering series is necessary (in difference,
e.g., to  $\pi d$ scattering, where a perturbative approach is 
justified). Such re-summation, however, can be carried out
analytically only in the static
limit. Bearing in mind that $M_K/m_N\simeq 0.5$, it is legitimate to ask, how
should the recoil corrections be systematically taken into account in the above
non-perturbative scheme.

\item[ii)]
Numerical estimates in the potential models 
show that the size of the recoil correction is moderate,
despite the fact that the kaon is quite heavy. One has to understand this 
observation on the basis of robust theoretical arguments.

\item[iii)]
The effective low-energy theory of QCD is ChPT. Since experiments
probe predictions of QCD, the theoretical framework used to
analyze the data, should be chosen accordingly. This means, e.g., that the
nucleon-nucleon and kaon-nucleon interactions should be described on the basis of effective
chiral Lagrangians. Because of  non-perturbative character of  $NN$ and $\bar K N$ interactions,
it could be however prohibitively complicated to try to describe
the three-body dynamics of the $\bar Kd$ system directly in ChPT. In 
Ref.~\cite{Baru:2009tx} we have formulated an alternative approach, based on the
use of the non-relativistic (non-local) effective Lagrangian, which   have
 the (chiral) two- and three-body potentials as an input,  see   Sec.\ref{sec:ansatz} for  more details.  Calculations of the $\bar Kd$ scattering
length can be systematically done within this approach.

\item[iv)]
The presence of the sub-threshold $\Lambda\,(1405)$ resonance in  $\bar KN$
scattering does not only lead to large scattering length. It also causes
a rapid variation of the amplitudes near threshold.  Therefore a  systematic study of  the (potentially large) effective-range 
corrections  is required.
 
\item[v)]
In the effective field theory approach, the three-body forces are necessarily
present and guarantee  independence of physical observables on the 
renormalization scale. Consequently,   a reliable  estimate of this contribution to the $\bar Kd$ scattering length
 is needed,  {see  Refs.\cite{Baru:2009tx,Baru:2012iv}  for further  discussions  of this subject.}  

\end{itemize}

In the present paper, we concentrate on the study of the recoil effect and
leave other issues for future publications. In particular, our main aim is to
formulate a procedure for including the recoil corrections {\it perturbatively}
 into the multiple-scattering series, in which the static interactions are 
summed up {\it to all orders}. 

The paper is organized as follows. In section~\ref{sec:ansatz} we discuss the
general scheme of including the recoil corrections into the multiple-scattering
series and discuss the counting scheme for such corrections. In section~\ref{sec:theory},  we 
derive   the  formulae  for the single insertion of the  recoil correction  in the multiple-scattering series. 
Explicit 
derivation of the formulae for  $\bar Kd$ scattering is given
in section~\ref{sec:theory}, where the numerical
results for one ``recoil insertion'' are also shown.
 In section~\ref{sec:xi1}, we discuss the expansion of a single recoil insertion in the  (non-integer) powers 
of the parameter $\xi=M_K/m_N$, along the similar lines to Ref.~\cite{Baru:2009tx},  and demonstrate the convergence of this expansion. 
In section~\ref{sec:numerics}, the numerical results for  the first-order recoil correction to the $\bar Kd$ scattering length are presented. 
Furthermore,  we  also calculate the boundaries  for  the antikaon-deuteron scattering length  using various $\bar K N$  scattering lengths existing in the literature 
as an input.    The section~\ref{sec:concl} contains our conclusions.

\section{Theoretical framework}\label{sec:ansatz}

The starting point of  the analysis is the generalized Deser-type relations between 
 the energy shift/width of the $1s$ level of the kaonic 
hydrogen ($\Delta E_{1s}, \Gamma_{1s}$) and kaonic deuterium 
($\Delta E_{1s}^d, \Gamma_{1s}^d$) and
 the pertinent $\bar KN$ and $\bar K d$ 
threshold amplitudes. These relations are well known in the 
literature (see, e.g., 
Refs.~\cite{Deser:1954vq,Meissner:2004jr,Meissner:2006gx,physrep}) and, 
at  order $O(\alpha)$ in isospin breaking, take the form
\begin{align}
&\Delta E_{1s}-i\Gamma_{1s}/2= -2\alpha^3\mu^2\,a_p(1-2\mu\alpha (\ln \alpha-1)a_p)\,,\\[2mm]
&\Delta E_{1s}^d-i\Gamma_{1s}^d/2= -2\alpha^3\mu_d^2\,{\cal A}(1-2\mu_d\alpha (\ln \alpha-1){\cal A})\,,\nn
\end{align}
where $\mu$ and $\mu_d$ are the reduced masses, and
$a_p$ and ${\cal A}$ are the threshold amplitudes of the
$\bar KN$ and $\bar K d$ systems, respectively. Moreover,
it can be shown (see Ref.~\cite{Baru:2009tx})
that  all logarithmically enhanced terms  of type $\alpha^{n}\ln^n\alpha$ 
for $n=1,2, \ldots$ can be re-summed to all orders, which leads to  
the replacement $1-2\mu\alpha (\ln \alpha-1)a_p\to (1+2\mu\alpha (\ln
\alpha-1)a_p)^{-1}$ and, similarly, to the replacement 
$1-2\mu_d\alpha (\ln \alpha-1){\cal A}\to (1+2\mu_d\alpha (\ln \alpha-1){\cal A})^{-1}$.  This procedure 
substantially improves the agreement between the modified Deser  formulae 
and  the explicit solutions within the framework of the potential models both 
for the kaonic hydrogen and kaonic deuterium\footnote{We thank A. Ciepl\'y for the
advanced communication of his numerical results to us.}. 

As seen from the above formulae, the measurement of the energy shift and width
of the kaonic hydrogen allows one to directly determine the (complex)
$K^-p$ scattering length $a_p$ that fixes two of four relations on the S-wave $\bar KN$
scattering lengths $a_0$ and $a_1$. The remaining two relations must be provided
from the eventually measured value of ${\cal A}$. 
A systematic expansion of the antikaon-deuteron scattering length in terms
of the $\bar KN$ scattering lengths (and higher-order effective-range expansion
parameters) naturally emerges through 
 the perturbative expansion of the threshold amplitude ${\cal A}$  in the non-relativistic effective field theories. 
The resulting scattering length is written in a form of a sum of all   six-point $\bar K NN \to \bar K NN$
diagrams, folded by the deuteron wave functions, and effective couplings in these diagrams  in a particular renormalization scheme   can be expressed  
in  terms of  the $\bar KN$ scattering lengths, effective radii and so on.
 However, due to the large values of the $\bar KN$ scattering lengths, the resulting 
multiple-scattering series for the six-point function does not converge. 
The only case when the multiple-scattering series can be algebraically
re-summed to all orders, is the so-called static approximation
$m_N\to\infty$. In this limit, the answer is explicitly written 
as~\cite{Kamalov,Meissner:2006gx}\footnote{In the kinematical prefactor, which appears in the r.h.s. of Eq.~(\ref{eq:static}), 
we have neglected the deuteron binding energy as compared to the nucleon mass.}
\eq\label{eq:static}
{\cal A}_{\sf st}=\frac1{1+\xi/2}\,
\int d^3{\bf r}\,|\Psi({\bf r})|^2\,
\frac{\tilde a_n+\tilde a_p+(2\tilde a_p\tilde a_n-b_x^2)/r
-2b_x^2\tilde a_n/r^2}{1-\tilde a_p\tilde a_n/r^2+ b_x^2\tilde a_n/r^3}\, .
\en
Here, $\Psi(r)$ stands for the wave function of the deuteron, normalized to 
unity. In momentum space, it obeys the equation
\eq
(\gamma^2+{\bf p}^2)\Psi({\bf p})=
\frac{1}{4m_N}\int\frac{d^3{\bf q}}{(2\pi)^3}V_{NN}({\bf p},{\bf q})\Psi({\bf q})\, ,\label{SGE}
\en
where $\gamma^2=m_N\varepsilon_d$ is the bound-state momentum in the deuteron 
while $\varepsilon_d$ denotes the binding energy. Further,
\eq\label{eq:ab}
\tilde a_{p,n,x,u}=(1+\xi)\,a_{p,n,x,u}\, ,\quad\quad
b_x^2=\tilde a_x^2/(1+\tilde a_u/r)\, ,
\en
where $a_{p,n,x,u}$ are the $\bar KN$ scattering lengths in the different physical channels 
$K^-p\to K^-p,~K^-n\to K^-n,~K^-p\to\bar K^0n$ and $\bar K^0n\to\bar K^0n$, respectively. In the isospin symmetry limit,
$a_p=a_u=(a_0+a_1)/2\, ,~
a_n=a_1\, ,~
a_x=(a_1-a_0)/2\, .$
{\it Numerically,}
in most cases, Eq.~(\ref{eq:static}) is a rather good approximation 
to the exact solution in spite of the 
relatively large kaon mass. 
In this paper, we demonstrate  that the recoil
corrections can  be calculated perturbatively, order by order  even though the
original multiple-scattering series (MSS)  is inherently non-perturbative. Below, we
describe, how this goal can be achieved.

An appropriate framework to this problem is the one based on the 
non-relativistic effective Lagrangians. We have described this 
framework in detail in Ref.~\cite{Baru:2009tx}. Here we only briefly mention
the crucial properties of the approach, focusing on the aspects which are new.

The approach relies on the existence of two distinct momentum scales. The 
nucleon-nucleon and three-particle interactions are characterized by a low
scale (of the order of the pion mass)
and are described by non-local, energy-independent
 potentials $V_{NN}({\bf p},{\bf q})$
and $V_3({\bf p}_1,{\bf p}_2,{\bf p}_3;{\bf q}_1,{\bf q}_2,{\bf q}_3)$, 
respectively. 
It is understood that, ultimately, these potentials should be taken
from 
chiral effective field theory in the two-nucleon sector. In the actual calculations
carried out in this paper we shall, however, use phenomenological nucleon-nucleon forces
which yield results similar to the chiral potentials and are easier to handle.
On the contrary, $\bar KN$ interactions are characterized
by a heavier scale (of the order of the mass of the $\rho,\omega,\cdots$
resonances). We shall describe these interaction by a tower of local
terms in the Lagrangian with zero, two,$\ldots$ space derivatives. The 
couplings emerging
in these terms are expressed through the $\bar KN$ scattering lengths, 
effective radii and so on in a standard manner. For this reason, a 
{\it perturbative} expansion in such an effective theory 
automatically yields the {\it multiple-scattering series}, known from the potential
scattering framework. The leading contribution to this series is provided 
by  the term with no derivatives which after re-summation yields Eq.~\eqref{eq:static}.
 
\begin{figure}[t]
\begin{center}
\includegraphics[width=12.cm]{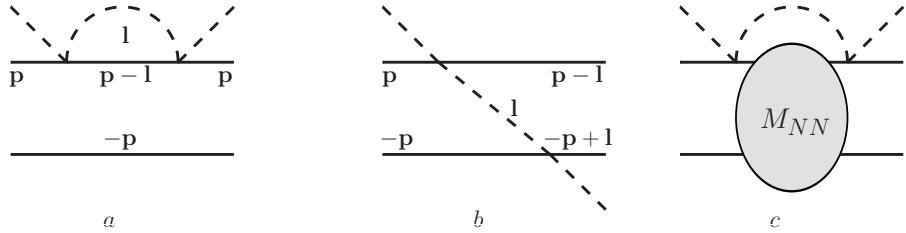}
\end{center}
\caption{The types of the diagrams emerging in the multiple-scattering
series: a) the self-energy type diagram; b) the exchange diagram;
c) any number of the ``potential exchanges'' between two nucleons in the intermediate state.
}
\label{fig:types} 
\end{figure}

A generic term in the multiple-scattering expansion contains the diagrams
in which the kaons are exchanged between two nucleons as well as kaons
hopping on the same nucleon (the self-energy-type diagrams), see 
Fig.~\ref{fig:types}. 
Since  $NN$ interactions are non-perturbative,  they  have to be included  to all orders. 
This is  normally done  by   solving  the Lippmann-Schwinger type equations which yield the 
$NN$ amplitudes at low-energies.  The $NN$ amplitude in its  turn  is  to be included in each intermediate state  of  the $\bar KNN-\bar KNN$ 
Feynman diagrams,  see e.g. diagram $c$ in Fig.\ref{fig:types}.
 
The renormalization prescription in the $\bar KN$ sector should guarantee that
the self-energy loop shown in Fig.~\ref{fig:types}a vanishes at threshold in the
$\bar KN$ center of mass frame.  Indeed,  the  self-energy loop in the two-body sector is  already  included as  a  part  of the relation between    the  effective coupling in the Lagrangian and 
the scattering length.  Therefore in the  calculation of  the  $\bar KNN-\bar KNN$ 
Feynman diagrams  where  the $\bar KN$ scattering lengths are  used  in the vertices  only that part  of the loop  survives  which appears  due to the  presence of  the 3-body  dynamics.      
The procedure  was described in detail, e.g.,  in Ref.~\cite{Baru:2004kw}  using $\pi N$ scattering as  an example. 
Using time-ordered perturbation theory, the expression for the self-energy diagram is given in the form
\eq
J_{S.E.}({\bf p}^2,E)&=&\int_{\sf reg}\frac{d^3{\bf l}}{(2\pi)^3}\, 
\frac{1}{{{\bf l}^2}/{2M_K}+{({\bf p}-{\bf l})^2}/{2m_N}+
{{\bf p}^2}/{2m_N}-E-i0}
\nonumber\\[2mm]
&=&\int_{\sf reg}\frac{d^3{\bf l}}{(2\pi)^3}\, 
\frac{1}{{{\bf l}^2}/{2\mu}+{{\bf p}^2}/{2(m_N+M_K)}+
{{\bf p}^2}/{2m_N}-E-i0}\, ,
\en
where the subscript {\footnotesize {\sf ``reg''}} stands for some (unspecified) ultraviolet regularization prescription, and $E$ stands for  the energy  with respect to the 
$\bar K d$ threshold.
In order to obtain the second expression, a shift of the integration variable
${\bf l}\to{\bf l}+\xi/(1+\xi)\,{\bf p}$ was necessary.
Consequently, the renormalized self-energy  (after  subtraction of the two-body term)  takes the form
\eq\label{SE}
J_{S.E.}^r({\bf p}^2,E)=\int
\frac{d^3{\bf l}}{(2\pi)^3}\,\biggl\{ 
\frac{1}{{{\bf l}^2}/{2\mu}+{{\bf p}^2}/{2(m_N+M_K)}+
{{\bf p}^2}/{2m_N}-E-i0}
-\frac{1}{{{\bf l}^2}/{2\mu}}\biggr\}\, .
\en
This expression indeed vanishes in the two-particle center of mass frame  at threshold (${\bf p}=0, E=0$).  On the other hand,  when this loop
is embedded in  the nuclear (deuteron)  environment,   the  energy  becomes  $E=-\varepsilon_d$,  and the nucleon momentum $\bf p$  does not vanish  anymore.

An appropriate parameter to establish the counting scheme for the 
recoil corrections is $\xi=M_K/m_N$. It can be easily seen that,     in the vicinity of the  static limit  ($\xi\to 0$) the renormalized self-energy graph vanishes as
$O(\xi^{1/2})$. Further, it can be also checked that  the operator with $NN$ interactions in the intermediate state, see Fig.~\ref{fig:types}c,
also leads to the contributions   suppressed as $O(\xi^{1/2})$.
Consequently, in the static limit, only diagrams of the type shown
in Fig.~\ref{fig:types}b survive. These are
the diagrams where the kaon is exchanged
between two nucleons. In this limit and assuming, in addition, $\varepsilon_d\to 0$,
the three-particle $\bar K NN$ propagator
in the exchange diagram simplifies to
\eq
g\doteq\frac{1}{{{\bf l}^2}/{2M_K}+{({\bf p}-{\bf l})^2}/{2m_N}+
{{\bf p}^2}/{2m_N}+\varepsilon_d}
\to \frac{1}{{{\bf l}^2}/{2M_K}}\,,
\en
so that the resulting series can be summed up to all orders. Taking the Fourier transform and folding the result
with the deuteron wave function, we finally arrive at the expression displayed
in Eq.~(\ref{eq:static}).

Away from the static limit, the propagator in the exchange 
diagram can be rewritten in the following form
\eq\label{green}
g=\frac{1}{{{\bf l}^2}/{2M_K}}+
\left\{\frac{1}{{{\bf l}^2}/{2M_K}+{({\bf p}-{\bf l})^2}/{2m_N}+
{{\bf p}^2}/{2m_N}+\varepsilon_d}
-\frac{1}{{{\bf l}^2}/{2M_K}}\right\}
\,\,\doteq\,\, g_{\sf st}+\Delta g\,, 
\en
where the  expression in the  curly  bracket, which is   defined as $\Delta g$, stands for the recoil correction.
The splitting shown above allows for the perturbative treatment of the 
recoil effect in the $\bar Kd$ scattering length. 
Schematically, the procedure can be described as follows.
Consider,  for example,   the diagrams of a type shown in Fig.~\ref{fig:types}b. 
Let $\tilde a$ denote a generic $\bar KN$ scattering length which is
related to the non-derivative coupling in the effective 
Lagrangian.  Then, Feynman diagrams  of type $b$   correspond to  consecutive scattering  of  the kaon off different nucleons  mediated  by  the  $\bar K NN$ Green functions
$g$  from Eq.~\eqref{green},  i.e.   $\tilde a g \tilde a  \cdots$.
The multiple-scattering series can be  therefore  written   as
\eq\label{recexp}
\tilde a+\tilde a g\tilde a+\tilde a g\tilde a g \tilde a+\cdots=
\left\{\tilde a+\tilde a^2g_{\sf st}+\cdots\right\}
+\left\{\tilde a+\tilde a^2g_{\sf st}+\cdots\right\}\,
(\Delta g)\,\left\{\tilde a+\tilde a^2g_{\sf st}+\cdots\right\}+\cdots\, .
\en
It is seen that the whole multiple-scattering
series can be rearranged so that the static contributions are re-summed to
all orders (the expression in the first  curly bracket in the r.h.s of Eq.~\eqref{recexp}), whereas the recoil corrections enter  perturbatively, in the form
of one, two, $\ldots$ ``recoil insertions'',  see the terms $\sim (\Delta g)^n$,  with $n=1, 2, \cdots$.   In Sec.~\ref{sec:xi1}
we shall prove that each insertion counts as  $O(\xi^{1/2})$  (or, in some cases, as  $O(\xi)$)  and, consequently, to carry out calculations at a given order in
the expansion parameter $\xi$, it suffices to consider a finite number of insertions.  

In the above equation, for  illustration, only the contribution from the 
exchange diagram was shown. The contributions from the diagrams shown in 
Fig.~\ref{fig:types}a and \ref{fig:types}c  are  parts  of   
 the recoil insertions, because they are absent in the static case.
We shall see later that they are also vanishing at $O(\xi^{1/2})$.

\begin{figure}[t]
\includegraphics[width=\linewidth]{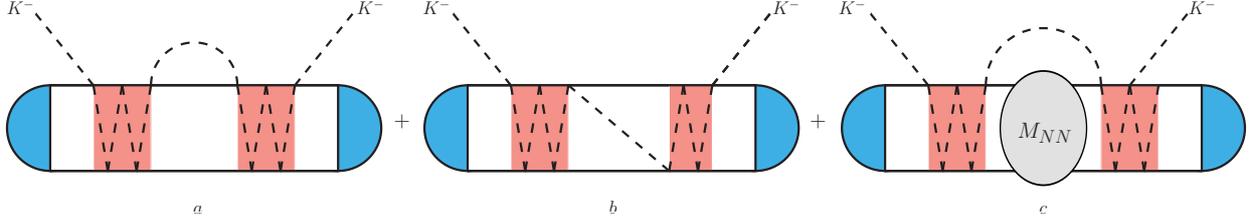}
\caption{Diagrams with one recoil insertion, where dashed (solid) lines 
denote the kaon (nucleon) lines, respectively. 
The whole diagram is folded by the deuteron wave functions (semicircles). 
The shaded boxes with kaon propagators symbolically denote the 
re-summed infinite series of kaon exchange graphs in the static limit, whereas
 the dashed lines outside of the boxes denote the retarded kaon 
propagators. The nucleon-nucleon amplitude is referred to as 
$M_{NN}$.}\label{oneins}
\end{figure}

Diagrams  with one recoil    insertion, which contribute to 
 the $\bar Kd$ scattering length are shown explicitly in Fig.~\ref{oneins}.
The case with two and more insertions can be treated similarly  and  is  beyond  the scope of this work.

Up to now, to the best of our knowledge, the recoil insertion has
been treated in the literature  only within a perturbative EFT
framework (see, e.g.,~\cite{Baru:2004kw,Baru:2009tx}). In the present paper,  we present  the perturbative  evaluation of the recoil insertions while 
 the static diagrams are summed up to all orders. 
Thus, the convergence of the multiple-scattering series is not assumed, and does 
not hold, in general.

\section{One recoil insertion}
\label{sec:theory}

In this section, we shall present compact explicit expressions for the 
recoil insertions in the multiple scattering series.  
As discussed in the previous section, the antikaon-deuteron scattering length can be written as 
\eq
{\cal A}={\cal A}_{\sf st}+{\cal A}^{(1)}+{\cal A}^{(2)}+\ldots, 
\en
where the individual terms correspond to the zero, one,  two, $\ldots$ 
recoil insertions. The lowest order  amplitude  corresponding to  the  static  limit  was  defined  by  Eq.~\eqref{eq:static}. 
 Next, consider one  recoil insertion, 
as   illustrated by the diagrams in Fig.~\ref{oneins}. 
 At  this order, one has the diagrams $a$ and $b$,
  where  the  kaon scatters either on the same nucleon or on different
nucleons,
as  well  as the diagram $c$  which  takes into account the intermediate  $NN$
interactions to all orders. Consequently,
\eq
{\cal A}^{(1)}={\cal A}^{(a)}+{\cal A}^{(b)}+{\cal A}^{(c)}\, ,
\en
where
\eq\label{RaRbRc}
{\cal A}^{(a)}&=&\frac1{1+\xi/2}\,
\int \frac{d^3{\bf p}d^3{\bf l}}{(2\pi)^6}\, \Gr({\bf p},{\bf l})
 \left[ 
 \Phi^2_p({\bf p+l/2})+\Phi^2_n({\bf p+l/2})+\Phi^2_x({\bf p+l/2})
 \right]\, ,
\nonumber\\[2mm]
{\cal A}^{(b)}&=&\frac1{1+\xi/2}\,
\int \frac{d^3{\bf p}d^3{\bf l}}{(2\pi)^6}\,
\biggl(G({\bf p},{\bf l}) -   G_{\rm st}({\bf l})\biggr) \\
&~& \times \left[
 \Phi_p ({\bf p+l/2} )\Phi_n ({\bf p-l/2} )+\Phi_n({\bf p+l/2})\Phi_p({\bf p-l/2})
 -\Phi_x({\bf p+l/2})\Phi_x({\bf p-l/2})  
 \right]\,,
\nonumber\\[2mm]\nonumber
{\cal A}^{(c)}&=&\frac{1}{1+\xi/2}\,
\int \frac{d^3{\bf p}d^3{\bf l}d^3{\bf q}}{(2\pi)^9}\,
 G({\bf p},{\bf l})\,  \biggl(\frac{\xi}{8\pi m_N}\,M_{NN}({\bf p},{\bf q},{\bf l})\biggr)\,  G({\bf q},{\bf l})\,
\\\nonumber
&~& \times\left[ \Phi_p({\bf p+l/2}) + \Phi_n({\bf p+l/2}) \right] \, 
\left[ \Phi_p({\bf q+l/2}) + \Phi_n ({\bf q+l/2}) \right]
 \,,
\en
and the different Green functions  
corresponding to the intermediate $\bar K NN$  state  read 
\eq\label{Green}
G_{\rm st}({\bf l})&=&\frac{4\pi}{{\bf l}^2}\, ,\quad\quad
G({\bf p},{\bf l})=\frac{4\pi}{{\bf l}^2(1+\xi/2)+2\xi ({\bf p}^2+\gamma^2)}\, ,\quad\quad
 G_{\rm r}({\bf p},{\bf l})=G({\bf p},{\bf l})-\frac{1}{1+\xi}G_{\rm st}({\bf l})\, .
\en
The nucleon-nucleon amplitude $M_{NN}$ which enters  ${\cal A}^{(c)}$ is 
determined from a solution of the Lippmann-Schwinger equation for a given 
two-nucleon potential $V_{NN}$
\eq
M_{NN}({\bf p},{\bf q},{\bf l})=V_{NN}({\bf p},{\bf q})
+\frac{\xi}{2m_N}\int \frac{d^3{\bf k}}{(2\pi)^3}\,
\frac{V_{NN}({\bf p},{\bf k})M_{NN}({\bf k},{\bf q},{\bf l})}
{{\bf l}^2(1+\xi/2)+2\xi ({\bf k}^2+\gamma^2)}\, ,
\en
where we used explicitly that the  energy relevant  for the study is  $E=-\varepsilon_d - l^2(1+\xi/2)/2m_K $.
Further, the quantity  $\Phi_i({\bf q})$  $(i=p,n,x)$ represents  the convolution of   the deuteron wave  function $\Psi(r)$ with the  re-summed  static  amplitudes~$A_i$  
\eq\label{phi}
\Phi_i({\bf q}) = \int d^3 {\bf r} \,  e^{i \bf q r} \,  \Psi(r)  \, A_i(r).
\en
The amplitudes  $A_i (r)$ obey  the  system of algebraic 
equations  in the $r$-space (see e.g.~\cite{Kamalov})
\eq\nonumber
A_p &=& \tilde a_p + \frac{\tilde a_p}{r} A_n -   \frac{\tilde a_x}{r} A_x\,,  \\ \nonumber
 A_n &=& \tilde a_n + \frac{\tilde a_n}{r} A_p\,,   \\
 A_x &=& \tilde a_x + \frac{\tilde a_x}{r} A_n -   \frac{\tilde a_u}{r} A_x\,, 
\en
where   $A_p$,  $A_n$ and $A_x$  are the amplitudes,
 in which the last (or the first)  interaction  of $K^-$ takes place on 
  the  proton ($K^- p\to K^- p$),  on the neutron  ($K^- n\to K^- n$) 
or is of the  charge-exchange-type ($K^- p\to \bar K^0 n$).  
The solution of this system yields  the required  amplitudes
\eq
A_p &=& \frac{\displaystyle \tilde a_p + \frac{\tilde a_p\tilde a_n -b_x^2}{r} - \frac{\tilde  a_n b_x^2}{r^2} } {\displaystyle
1 - \frac{\tilde a_p\tilde a_n}{r^2} + \frac{\tilde  a_n b_x^2}{r^3} }\,,      \quad \quad
 A_n = \frac{\displaystyle \tilde a_n + \frac{\tilde a_p\tilde a_n}{r} - \frac{\tilde  a_n b_x^2}{r^2} } {\displaystyle
1 - \frac{\tilde a_p\tilde a_n}{r^2} + \frac{\tilde  a_n b_x^2}{r^3} }\,, 
\quad\quad
 A_x = \frac{\displaystyle \tilde a_x \left(1 + \frac{\tilde   a_n }{r}   \right)  } {\displaystyle
1 - \frac{\tilde a_p\tilde a_n}{r^2} + \frac{\tilde  a_n b_x^2}{r^3} }\, \frac{\displaystyle1}{\displaystyle 1 + \frac{\tilde   a_u }{r} }\,.
\en
Here, the quantity $b_x$ is defined in Eq.~\eqref{eq:ab}.  It is   straightforward to see that if   
one  does the  replacement  $A_i \to \tilde a_i$,    the expressions  \eqref{RaRbRc}  transform  
exactly to  those  derived  in Ref.~\cite{Baru:2009tx} for  the   double  scattering diagrams.
Note  that  the symmetric combination $A_p+A_n$  represents  the isoscalar contribution  to the  $\bar KNN -\bar KNN $  operator  which  does not change the isospin state  of the $NN$-pair.  Being convoluted with the deuteron wave functions in the initial and  final states,  it  yields  the static amplitude \eqref{eq:static} of $\bar K d$ scattering. 
Similarly,   the  antisymmetric combination of the amplitudes  $A_p-A_n$ as well as  the charge exchange  amplitude $A_x$ represent the 
isovector  contributions  to the  $\bar KNN -\bar KNN $  operator.   When applied to the  initial deuteron state, these operators   give rise to   the isospin-1 $NN$-pair.  
Following Refs.~\cite{Baru:2009tx, Baru:2004kw},  we therefore decompose  the  results \eqref{RaRbRc}  
 into the contributions corresponding to the two-nucleon intermediate states with a certain isospin 
 \eq\label{Aiso}
 {\cal A}^{(1)}={\cal A}^{(a)}+{\cal A}^{(b)}+{\cal A}^{(c)} =  \biggl\{{\cal A}_{1} +   \Delta {\cal A}_{{\sf st},1}\biggr\} + 
 \biggl\{{\cal A}_{0} +   {\cal A}^{(c)}  +\Delta {\cal A}_{{\sf st},0}\biggr\}\, , 
\en
where  the expression in the first  curly bracket in the r.h.s. of Eq. \eqref{Aiso}  corresponds to the $I=1$ $NN$  
intermediate state while  that  in the second bracket corresponds to $I=0$.  
In order to obtain Eq. \eqref{Aiso},   the  expressions  for  ${\cal A}^{(a)}$ and ${\cal A}^{(b)}$  were  rewritten  using  the following rules:
\begin{enumerate}
\item  The  combinations of the functions  $\Phi_i$  (cf. Eq.~\eqref{phi})   which enter  the expressions for  ${\cal A}^{(a)}$ and ${\cal A}^{(b)}$ can be decomposed  
into the isoscalar and isovector  parts     using the  relations
\eq\nonumber
 \Phi^2_{p1}+\Phi^2_{n1}+\Phi^2_{x1}&=&   \frac12 \left(\Phi_{p1}+\Phi_{n1}\right ) \left(\Phi_{p1}+\Phi_{n1}\right )+
  \frac12 \left(\Phi_{p1}-\Phi_{n1}\right ) \left(\Phi_{p1}-\Phi_{n1}\right )
   +\Phi_{x1}^2 \,,\\
    \Phi_{p1}\Phi_{n2}+ \Phi_{n1}\Phi_{p2}-\Phi_{x1}  \Phi_{x2}&= &  \frac12 \left(\Phi_{p1}+\Phi_{n1}\right ) \left(\Phi_{p2}+\Phi_{n2}\right )-
  \frac12 \left(\Phi_{p1}-\Phi_{n1}\right ) \left(\Phi_{p2}-\Phi_{n2}\right )
   -\Phi_{x1}\Phi_{x2}.
\nonumber
\en
where  the second subscript indicates that the  functions $\Phi$ in 
Eq.~\eqref{RaRbRc} depend on two  different  momenta  $\bf   l_1={\bf p+l/2}$  and $\bf  l_2={\bf p-l/2}$, respectively. 
Since  the $\Phi_i$  are proportional to the $A_i$,  the  first  term in the r.h.s  of these equations  contributes to   $I=0$ $ NN$ intermediate state  while  the others  correspond to $I=1$.
Using these relations one determines the  resulting   $I=0$  and  $I=1$ contributions in the curly brackets  of  Eq.~\eqref{Aiso}. 
\item  It appears  useful to  further decompose the  net $I=0$  and
  $I=1$ contributions in such a way  that 
the  terms   ${\cal A}_0$   and ${\cal A}_1$  contain  
the same renormalized $\bar K NN$ Green  function  $G_{\rm r}$   (see  Eq.~\eqref{Green}). 
The trivial  corrections  $\Delta {\cal A}_{{\sf st},0}$  and 
$\Delta {\cal A}_{{\sf st},1}$, which can be expanded into the
 integer powers of $\xi$,   appear  as the result   of this  separation as well.
 \end{enumerate}
Thus, one gets
\eq\label{A01}
{\cal A}_0&=&\frac1{2}\frac1{1+\xi/2}\,
\int \frac{d^3{\bf p}d^3{\bf l}}{(2\pi)^6}\, 
\left[ \Phi_p({\bf l_1})+\Phi_n({\bf l_1}) \right]\, \Gr({\bf p},{\bf l})
 \left[ 
 \Phi_p({\bf l_1})+\Phi_n({\bf l_1})+\Phi_p({\bf l_2})+\Phi_n({\bf l_2})
 \right]\, ,
\nonumber\\[2mm]
{\cal A}_{1}&=&\frac1{2}\frac1{1+\xi/2}\,
\int \frac{d^3{\bf p}d^3{\bf l}}{(2\pi)^6}\, 
 \Gr({\bf p},{\bf l})  \\ &~&\times
 \left[ \left(\Phi_p({\bf l_1})-\Phi_n({\bf l_1}) \right)\,
 \left(
 \Phi_p({\bf l_1})-\Phi_n({\bf l_1})-\Phi_p({\bf l_2})+\Phi_n({\bf l_2})
 \right)  +2\Phi_x({\bf l_1}) \left( \Phi_x({\bf l_1})-\Phi_x({\bf l_2})\right)
 \right]
 \, ,\nonumber
\en
and 
\eq\label{delA}\nonumber
 \Delta {\cal A}_{{\sf st},0}&=&-\,\frac{\xi}{2(1+\xi)(1+\xi/2) } \int d^3{\bf r} \frac{\Psi^2({\bf r})}{r} \left(  
 A_p(r)+A_n(r)   \right)^2\,, \\
  \Delta {\cal A}_{{\sf st},1}&=&\,\frac{\xi}{2(1+\xi)(1+\xi/2)  }
\int d^3{\bf r} \frac{\Psi^2({\bf r})}{r} \left(  
 \,(A_p(r)-A_n(r))^2+2A_x(r)^2 
 \right)\, .
\en
The expressions  written in the  isospin basis  allow  for a  very  clear  physical  interpretation of the  cancellations  between various individual recoil  corrections, as  will be  discussed in section~\ref{sec:cancel}.

\section{Perturbative expansion of the first-order correction in powers of $\xi$}
\label{sec:xi1}

\subsection{Formalism}

The expression of the single recoil insertion, which was given in 
the previous section, depends on the variable $\xi$ in a non-trivial way.
In order to establish systematic power-counting rules, it is therefore necessary
to perform an expansion of this expression in $\xi$.
Further, considering such an expansion
helps one to reveal the pattern of cancellations of the leading terms, which has
been discussed already in Ref.~\cite{Baru:2009tx}. Note that these cancellations
are crucial for discussing the convergence of the perturbative method,
which is described in the present paper.
Note also that the final {\it numerical} results presented in section~\ref{sec:numerics}
include the {\it full} expression of the one recoil insertion and do not 
rely on this expansion.

As already shown in Ref.~\cite{Baru:2009tx}, the $\bar K d$ scattering length
can be systematically expanded in the half-integer powers of $\xi$. As an illustrative example,  the expansion of
the  double  scattering diagrams was performed, and a relatively rapid
convergence  was  observed. In this  paper,  we  investigate  the  convergence of  this expansion 
for the  case  of multiple  scattering  series  with one recoil insertion. 
 
Below,  we shall use the
uniform expansion method, see Refs.~\cite{Beneke:1997zp,Mohr:2005pv}.
In this method, different momentum regions are identified,
 according to the scales appearing in the problem. The integrand
is expanded in each region before performing the integration, and the results are summed up.
In our case, the relevant regions are defined by two momentum  scales  in  the  full  $\bar K NN$  Green function~\eqref{Green}.
 One therefore  finds  three  relevant integration  regions:

\subsubsection*{1) low-momentum region:}

\noindent
The internal kaon momentum $l$ is much smaller than the typical 
deuteron scale $p\sim \langle 1/r\rangle \sim m_{\pi}$, 
but still of the same order of magnitude as  $\sqrt{\xi}p$. 
In this region, the Green function $G({\bf p},{\bf l})$ ``feels`` the presence 
of the nearby  3-body  singularity.
Meanwhile,    all other quantities,  such as  the  functions  
$\Phi_i ({\bf p \pm l/2})$, can be  systematically expanded,
  assuming  $l\ll p$.  
Given that  $l\sim \sqrt{\xi}p$, non-integer orders of $\xi$ emerge after the integration.

 \subsubsection*{2) high-momentum region:}

\noindent
 In this region, $l\sim p$ and all retarded terms in 
$G({\bf p},{\bf l})$  are suppressed 
 by $\xi$. The direct  expansion in $\xi$  leads to the  
appearance of integer orders in the expansion.

 \subsubsection*{3) intermediate region:} 

\noindent
The momenta in the  intermediate 
region obey $\sqrt{\xi}p\ll l\ll p$. The integrand in 
this region  can be obtained, 
 if one applies  the  low-momentum  expansion to the integrand  already 
expanded  in the high-momentum region or vice versa.  

\smallskip

The original  integrand ($I$) of the amplitude  can then be written as 
\eq
I= I_{\rm low} + I_{\rm high} - I_{\rm interm} \,.
\en
Evaluating the integral over the whole momentum range, we obtain  
the amplitudes at a given order in   $\xi$. 
Note  that the integrand, being expanded in the regions 1) and 2),  
contains   ultraviolet- as well as  infrared-divergent terms. However, these 
divergences cancel exactly with those that emerge from the intermediate region, rendering 
the result finite and independent of the regularization scheme.

\subsection{Leading-order  recoil correction}
\label{sec:cancel}

\begin{figure}[t]
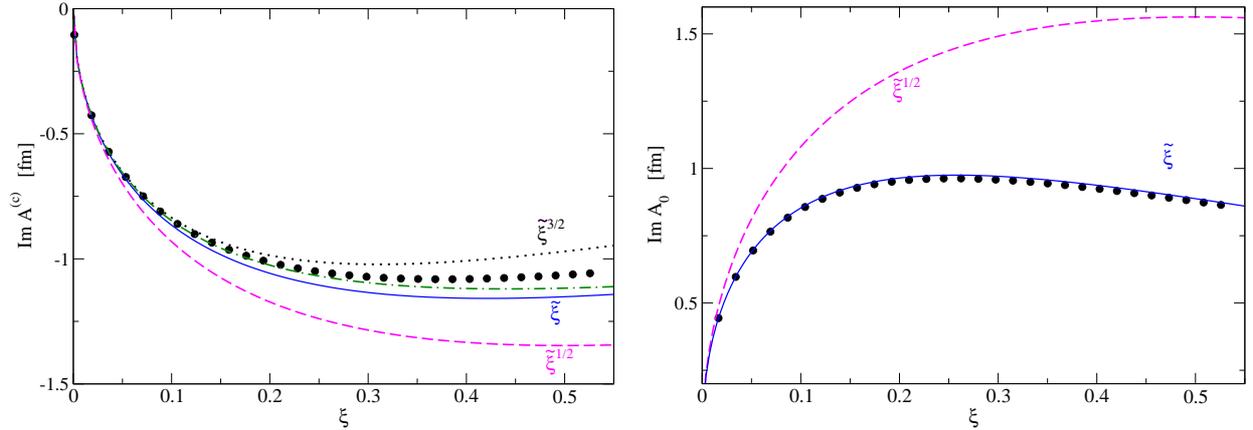

\includegraphics[width=0.505\linewidth]{ImANN_one_ins_xi_tilde.eps}
\includegraphics[width=0.48\linewidth]{ImA0_one_ins_xi_tilde.eps}
\caption{Expansion of the amplitude ${\cal A}_0$ and ${\cal A}^{(c)}$ in powers of $\tilde\xi=\xi/(1+\xi/2)$:  left panel -- ${\rm Im}\,{\cal A}_0$,  right panel  -- ${\rm Im}\,{\cal A}^{(c)}$.  Dashed, solid,  dotted, and dot-dashed lines  correspond  to  the expansion up to the order  $\tilde\xi^{1/2}$, $\tilde\xi$, $\tilde\xi^{3/2}$,  and $\tilde\xi^{2}$, respectively.
The result of the full calculation
is presented by the black dots. }\label{fig:exp0}
\end{figure}

It has been known for  many  years that, due to the emerging cancellations,
 the   recoil corrections   to   pion-deuteron scattering at threshold  are smaller than one would naively expect, see, 
e.g., Refs.~\cite{Faeldt, kolyb,Baru:1997xf}. 
 In the isovector channel, relevant for    $\pi d$ scattering, the mechanism behind  this cancellation  was identified   
in Ref.~\cite{Baru:2004kw}, see also  Refs.~\cite{JOB,Lensky_gam}  for  further  discussions.   In  Ref.~\cite{Baru:2009tx},  the arguments were summarized and generalized  to the  isoscalar 
channel  which  is  of importance for $\bar Kd$ scattering.  In all these  studies, however,  the recoil effect was studied  for the  double-scattering process only  that is fully 
justified for   $\pi d$ scattering  but   does not suffice  for  $\bar Kd$ scattering.   Here  we discuss,  what happens  with the  leading  $O(\xi^{1/2})$  correction in the presence of  the static 
multiple-scattering  ladders.  If  this correction survives,  it  would  give  
an estimated  70\%  correction  to  $\cal A$,  in contradiction  to a  relatively  good  agreement between the static  MSS  and Faddeev calculations.

The origin of the cancellation  of  the leading  isovector  correction in
 the quantity ${\cal A}_{1}$ from Eq.~\eqref{A01}
  lies in  the  fact that  the $NN$ pair with total isospin $I=1$
 must 
have  orbital momentum  $L=1$,  as  a  consequence of the Pauli  selection rules.  Hence,  the  diagram  $c$  in Fig. \ref{oneins},   which accounts  for  
the S-wave  $NN$  interaction, 
does not  contribute in this case.  Meanwhile,   the diagrams $a$  and $b$  do   produce   $O(\xi^{1/2})$  corrections individually.
 Those,  however, correspond to the  $NN$ states     in  
the S-wave  and thus  naturally
cancel in the sum.  This can be seen immediately
since, as argued above, all corrections with the non-integer powers of $\xi$
 emerge
  from  the low-momentum  region. 
Expanding then the integrand in Eq.~\eqref{A01} in the assumption 
 $l\ll p$,   one immediately finds that  the leading contribution to 
${\cal A}_{1}$ vanishes. 
 
On the contrary, in the isoscalar  case,  the  $NN$
  intermediate states can appear in the S-wave. This
 means that  all diagrams  in Fig. \ref{oneins} are relevant. 
 Furthermore, for the isoscalar  
$\bar K N$-interaction,     the  quantum numbers of  the intermediate $NN$
 pair  must  coincide  with those  of the deuteron, i.e.  the  $NN$ 
state  appears in the $^3 S_1$ partial wave.   It  was shown in 
Ref.~\cite{Baru:2009tx}  that  the sum of the  diagrams of  type shown in  
Fig. \ref{oneins}    gives the  full Green function of the 
$NN$ state,
 which should be orthogonal to  the bound-state deuteron wave function  due to the completeness condition. 
This was  the origin of cancellation of the isoscalar recoil  corrections  in  the  double-scattering  diagrams,  considered in Ref.~\cite{Baru:2009tx}.
The situation is, however,  different in the multiple-scattering case, since the  infinite  static  ladders (see the shaded blocks  in Fig. \ref{oneins})
separate  the  bound-state  and continuum wave  functions,
     providing  the  effect  of screening.  Therefore, the cancellation of  the isoscalar  $O(\xi^{1/2})$  recoil correction is  not exact,  although
it is  still largely present,
as   will be  shown below.
 
\subsection{Convergence of  the expansion in $\xi$}

\begin{figure}[t]
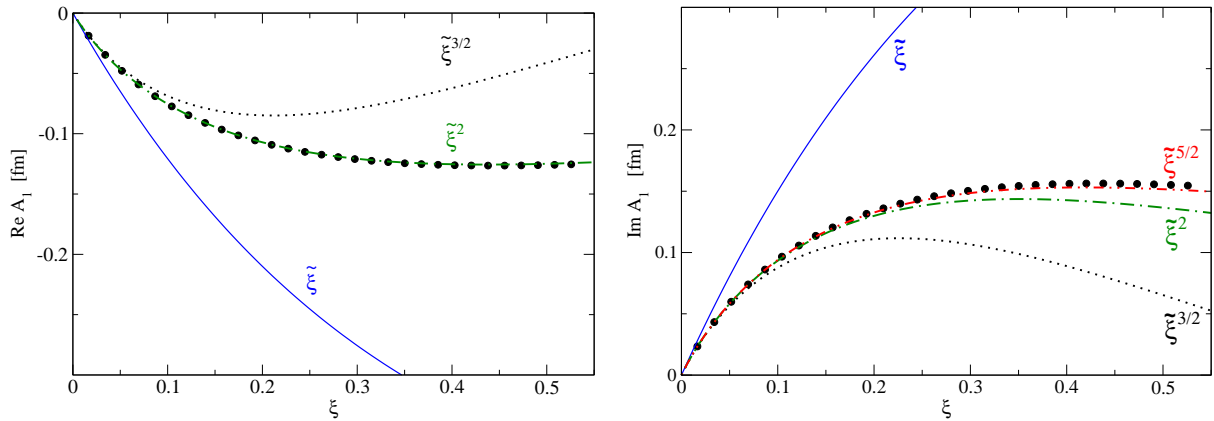

\includegraphics[width=0.47\linewidth]{ReA1_one_ins_xi_tilde.eps}\hspace*{.3cm}
\includegraphics[width=0.47\linewidth]{ImA1_one_ins_xi_tilde.eps}
\caption{Expansion of the amplitude ${\cal A}_1$ in powers of $\tilde\xi=\xi/(1+\xi/2)$:  left panel -- ${\rm Re}\,{\cal A}_1$,  right panel  -- ${\rm Im}\,{\cal A}_1$.  Solid,  dotted, dot-dashed and  double-dot-dashed 
lines  correspond  to  the expansion up to the order  $\tilde\xi$, $\tilde\xi^{3/2}$, $\tilde\xi^{2}$, $\tilde\xi^{5/2}$, respectively.
The result of the full calculation
is presented by the black dots.}\label{fig:exp1}
\end{figure}

In Fig. \ref{fig:exp0}   we  demonstrate the  convergence of  the expansion for the isoscalar  case.
The Hulth\'{e}n $NN$ potential is used in the calculations\footnote{This potential is regular at short distances (does not depend on the large mass scales). 
Consequently, there is no need for introducing a short-distance regularization that simplifies the calculation substantially.}.
Note further that we found it useful to introduce
a modified expansion parameter $\tilde\xi=\xi/(1+\xi/2)$. The convergence of the series in this new parameter improves substantially, see \ref{app:exp}  for the explicit expressions. 
 For illustrative purposes, we  present  the   results   for  ${\rm Im}\,{\cal A}_{0}$  and  ${\rm Im}\,{\cal A}^{(c)}$ separately,  both   dominated  by  the leading order $O( \tilde\xi^{1/2})$ terms. 
 Meanwhile,  these  corrections  appear with  different signs  and  thus largely  cancel  in the sum.  
It is interesting that  already  at  the order   $ \tilde{\xi}$,  the expanded  results  nicely agree  with  
 the exact ones.     Analogous results  for the  isovector  case  are  shown in Fig. \ref{fig:exp1}  for    ${\rm Re}\,{\cal A}_{1}$ (left panel)  and  ${\rm Im}\,{\cal A}_{1}$ (right panel).    
  In this case,  the expansion  starts  from the  order  $\tilde\xi$, in accordance with the  Pauli selection rules.  The expansion in half-integer
powers of $\tilde\xi$ for  the  isovector  case 
  converges a little slower  than for the isoscalar one, but still provides a  very good approximation to the unexpanded  results  already at the order $\tilde\xi^2$.   

\section{Numerical results for the $\bar Kd$ scattering length}
\label{sec:numerics}
\subsection{The role of recoil effects for kaonic deuterium}

We are now in the position to analyze the role of the recoil corrections in the 
$\bar Kd$ scattering length. The results of our calculations for one recoil
 insertion
are given in Table~\ref{tab:no-expansion}. As already mentioned, at this stage,
we use the phenomenological nucleon-nucleon separable potentials
instead of the more complicated potentials, constructed within the chiral EFT. 
In  particular, the  results of the calculations shown in 
Table~\ref{tab:no-expansion} are obtained for   the Hulth\'{e}n 
and PEST $NN$ potentials, see~\ref{app:potentials}. The  following 
values of the $\bar K N$ scattering lengths are used
as an input~\cite{Shevchenko:2011ce}:
\eq\label{KNinput}
a_0= (-1.62 + i0.78)~\mbox{fm}\, ,\quad  a_1= (0.18 + i0.68)~\mbox{fm}\, ,
\en

The following conclusions can be drawn, based on    the results
shown in Table~\ref{tab:no-expansion}:

\setlength\extrarowheight{1pt}
\begin{table}[t]
\begin{center}
\begin{tabular}{|c|l|ccc|}
\multicolumn{5}{c}{\textbf{Hulth\'{e}n potential}}\\
\hline
${\cal A}_{\sf st}$&\multicolumn{4}{c|}{$-1.49+i1.19 $}\\
\hline
\multirow{6}{*}{${\cal A}^{(1)}$}&${\cal A}_1$&&$-0.13+i0.16 $&\\
				 &$\Delta {\cal A}_{{\sf st},1}$&&$ +0.12-i0.20 $&\\
\cline{4-4}
				 & &&$ +0.00-i0.04 $&\\[2mm]
				 &$  {\cal A}_0$&&$-0.27+i0.87 $&\\
				 &$\Delta {\cal A}_{{\sf st},0}$&&$-0.12+i0.33 $&\\
				 &${\cal A}^{(c)}$&&$+0.35 -i1.06  $&\\
\cline{4-4}
				 & &&$-0.03+i0.13 $&\\[2mm]
\cline{2-5}
				 &Sum:~~~&&$-0.03+i0.09 $&\\
\hline\hline
${\cal A}_{\sf st}+{\cal A}^{(1)}$&\multicolumn{4}{c|}{$-1.52+i1.27 $}\\
\hline
\end{tabular}
\hspace{+1.5cm}
\begin{tabular}{|c|l|ccc|}
\multicolumn{5}{c}{\textbf{PEST potential}}\\
\hline
${\cal A}_{\sf st}$&\multicolumn{4}{c|}{$-1.55+i1.25 $}\\
\hline
\multirow{6}{*}{${\cal A}^{(1)}$}&${\cal A}_1$&&$-0.13+i0.18 $&\\
				 &$\Delta {\cal A}_{{\sf st},1}$&&$+0.13-i0.22$&\\
\cline{4-4}
				 & &&$+0.00-i0.04 $&\\[2mm]
				 &$  {\cal A}_0$&&$-0.29+i0.97 $&\\
				 &$\Delta {\cal A}_{{\sf st},0}$&&$-0.11+i0.34$&\\
				 &${\cal A}^{(c)}$&&$+0.36-i1.19 $&\\
\cline{4-4}
				 & &&$-0.04+i0.12 $&\\[2mm]
\cline{2-5}
				 &Sum:~~~&&$-0.04+i0.08 $&\\
\hline\hline
${\cal A}_{\sf st}+{\cal A}^{(1)}$&\multicolumn{4}{c|}{$-1.59+i1.32 $}\\
\hline
\end{tabular}
\end{center}
\caption{The recoil  corrections to the $\bar Kd$ scattering length
  from one insertion of  the  retarded block. All quantities are given
  in units of fm.}
\label{tab:no-expansion}
\end{table}

\begin{itemize}

\item[i)]
In the isoscalar channel,
 the individual contributions, which still contain the dominant $O(\tilde\xi^{1/2})$  term,  are very  large, especially the imaginary parts thereof. 
However,  they  undergo  significant  cancellations, 
yielding  only about  a 10\%  net correction 
 to  the imaginary part of the static  term.
 
 \item[ii)]
 The resulting isovector recoil  correction 
appears  to be even smaller  providing  only about  a 3\%  correction to the static term.  
Its smallness can be understood  from the exact  cancellation of  the  
$O(\tilde\xi^{1/2})$  term along with some additional  cancellations among
  higher-order  terms.  
This effect was already  discussed in Ref.~\cite{Baru:2009tx}, see also  Fig.~\ref{fig:exp1}.

\item[iii)]  The recoil contributions $\Delta {\cal A}_{{\sf st},0}$ and $\Delta {\cal A}_{{\sf st},1}$, which start to contribute at the order $\tilde\xi$, are still sizable. 
These are, however, largely canceled  by other contributions, pointing  at cancellations that emerge at this and  higher
orders in $\tilde\xi$. At this stage, the mechanism of such cancellations is unclear and needs to be addressed. 

\item[iv)]
The net  correction ${\cal A}^{(1)}$ which stems from one recoil insertion
 in the MS  diagrams  appears to be  quite small, of order
of  $\simeq 6-8\%$  of  ${\cal A}_{\sf st}$,
 despite the large value of $\xi$.
An  additional  suppression is  partly accounted  for  by   
cancellations of the isoscalar   and isovector recoil  corrections.

\item[v)]
Ref.~\cite{Shevchenko:2011ce} gives  the value of
the ${\bar Kd}$ scattering length, obtained by
the solution of the  Faddeev equations  with the  one-channel 
energy-independent optical potential.   The  parameters of  this potential
were  adjusted  to reproduce the  scattering  lengths \eqref{KNinput}.   
The  calculation, carried in Ref.~\cite{Shevchenko:2011ce}, yields
 ${\cal A}\simeq (-1.47 + 1.11i)~\mbox{fm}\, $, see the 
half-empty square in Fig. 12 in that paper. 
 
Although the input in  the two  calculations  might  still     
differ  in several aspects,  such as  different  $NN$   models,
 or  the use  of the off-shell form factors,  needed to regularize the
Faddeev calculations, 
it is  instructive to compare this  result  with  the  pertinent  results  
shown in   Table~\ref{tab:no-expansion}.
In particular,  one  can see  that  
 the differences between the static result and the solution of the  
Faddeev equation are, generally, of the same order as the full recoil 
correction ${\cal A}^{(1)}$.

\end{itemize}

Finally, few words should be said about the higher-order recoil corrections
(two or more recoil insertions). Albeit the results with one insertion
look encouraging, these do not completely guarantee the convergence of the
procedure, since the smallness of the first-order correction could still
follow from some peculiar cancellations. In this respect, according to
 the above discussion, the study of the corrections to 
the {\it isoscalar} amplitude, where the cancellation of
the leading  $O(\tilde\xi^{1/2})$ terms is not complete due to the presence of the static ladders, appears to be 
crucial.

\subsection{Prediction for  the   $\bar K d$  scattering length}

\begin{figure}[t]
\includegraphics[width=1.\linewidth]{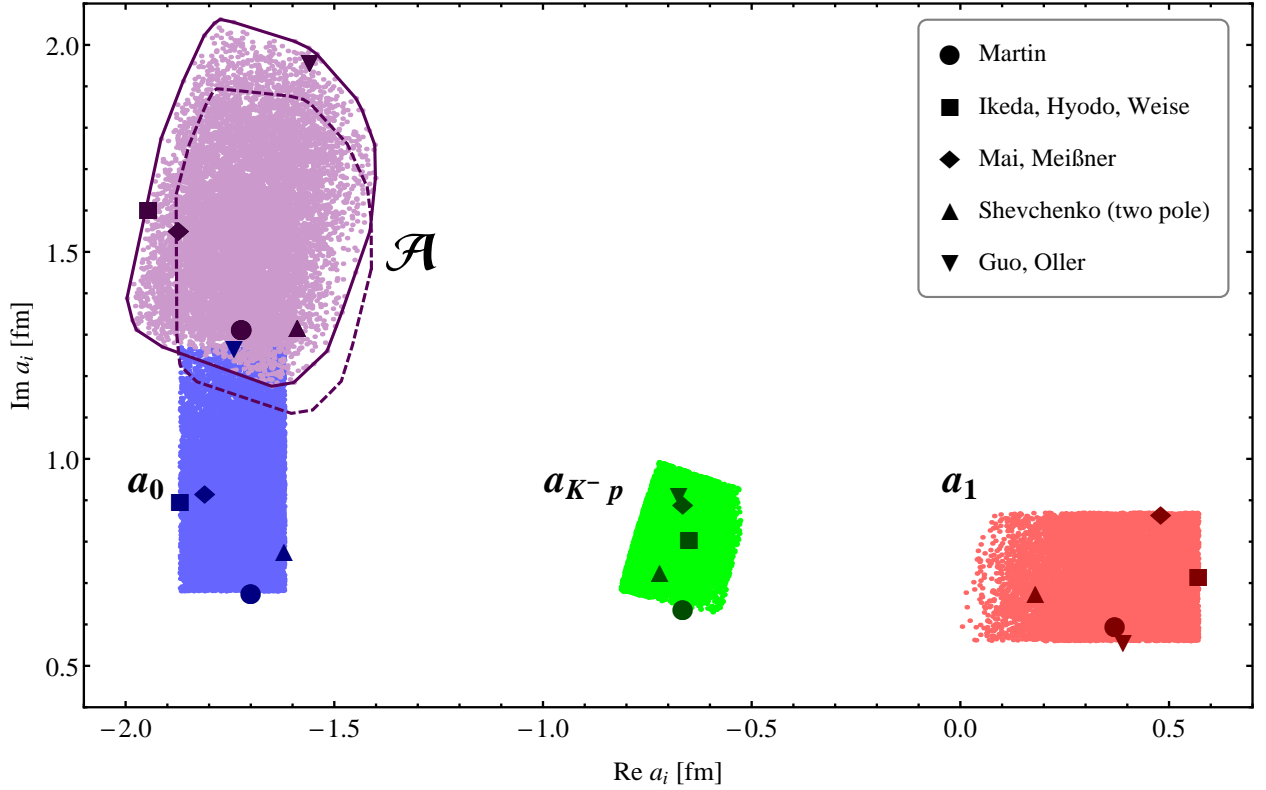} 
\caption{The antikaon-deuteron scattering
length ${\cal A}$ (purple dots) for given input values of the
antikaon-nucleon scattering lengths $a_0$ (blue) and $a_1$ (red). 
These lengths are chosen to be randomly distributed in the rectangular areas, restricted by results from the SIDDHARTA experiment ~\cite{Bazzi:2011zj}, see 
Eq.~\eqref{eq:rectangle}. The full purple line shows the
convex hull of all results for the $\bar Kd$ scattering length, 
whereas the purple dashed line is the
convex hull for the static results (not shown explicitly). }\label{Kd}
\end{figure}

In this paper, we developed an EFT framework which directly relates
 the $\bar KN$
scattering lengths to the ${\bar Kd}$ scattering length. Since the latter is
not yet measured in the experiment, we find it useful to predict in which
region this quantity could lie for various input values of the $\bar KN$
scattering lengths, known from the literature. This procedure allows us to test
the sensitivity of the final outcome to the input and, what is not less important,
the sensitivity of  the recoil correction to the same input.

We have adopted the following strategy. We assume that the $NN$ interaction
is described by the PEST potential~\cite{Zankel:1983zz} and examine
a broad variety of the models for the $\bar KN$ interaction. Some of the models
are listed in Table~\ref{tab:Kd}. For each of the models, using the quoted
values of the $\bar KN$ scattering lengths $a_0$ and $a_1$, we calculate the
 $\bar Kd$ scattering length in the static approximation, as well as including the
first-order  recoil correction. The results are presented in Fig.~\ref{Kd}.

Moreover, in order to minimize the dependence on the input, we randomly generate
 $\bar KN$ scattering lengths from the rectangular areas
\eq\label{eq:rectangle}
-1.87\leq {\rm Re}\,a_0\leq\,-1.62 ,\quad
+0.68\leq {\rm Im}\,a_0\leq\,+1.27 ,\quad
-0.06\leq {\rm Re}\,a_1\leq\,+0.57 ,\quad
+0.56\leq {\rm Im}\,a_1\leq\,+0.87 .
\en
These intervals include most values of $a_0$ and $a_1$, known in the literature. Further,
for each  random pair $a_0$ and $a_1$ we check that $a_p=(a_0+a_1)/2$ obeys the constraints\footnote{Here,
for consistency reasons, we have neglected the isospin-breaking effects in the quantity $a_p$.
These effects might become substantial, see Ref.~\cite{Meissner:2004jr}.}
imposed by the SIDDHARTA experiment~\cite{Bazzi:2011zj} and reject the
pairs for which this constraint
is not fulfilled.
Again, we calculate the $\bar Kd$ scattering length in the static 
approximation, 
as well as with first-order  recoil correction. The dotted and solid curves in 
 Fig.~\ref{Kd} show the convex hulls for the results of these calculations, determining
the region where the measured value of the $\bar Kd$ scattering length is expected to lie.

\begin{table}[t]
\begin{center}
 \begin{tabular}{|l|c|c|c|c|}
\hline
Reference & ~~~~$a_0$ [fm]  & ~~~~$a_1$ [fm]  & ${\cal A}_{\sf st}$ [fm]& ${\cal A}={\cal A}_{\sf st}+{\cal A}^{(1)}$ [fm] \\
\hline
Martin \cite{Martin:1980qe}			&$-1.70~+i0.68$	&$+0.37+i0.60$	
&$-1.65+i1.23$&$-1.72+i1.32$\\
Ikeda, Hyodo, Weise \cite{Ikeda:2012au}	&$-1.87+i0.90$	&$+0.57+i0.72$
&$-1.87+i1.48$&$-1.95+i1.61$\\
Mai, Mei\ss ner \cite{Mai:2012dt}		&$-1.81+i0.92$	&$+0.48+i0.87$	
&$-1.82+i1.45$&$-1.87+i1.56$\\
Shevchenko (two pole) \cite{Shevchenko:2012np}&$-1.62+i0.78$	&$+0.18+i0.68$	
&$-1.55+i1.25$&$-1.59+i1.32$\\
Guo, Oller \cite{Guo:2012vv}			&$-1.74+i1.27$	&$+0.39+i0.56$	
&$-1.58+i1.83$&$-1.56+i1.96$\\
Borasoy, Mei\ss ner, Ni\ss ler \cite{BMN}		&$-1.64+i0.75$	&$-0.06+i0.57$	
&$-1.49+i1.12$&$-1.53+i1.18$\\
\hline
\end{tabular}
\end{center}
\caption{The $\bar Kd$ scattering length for different input values of the $\bar KN$ scattering lengths.}
\label{tab:Kd}
\end{table}

Several important observations can be made from Table~\ref{tab:Kd} and Fig.~\ref{Kd}.
First of all, the $\bar Kd$ scattering length turns out to be quite sensitive
to the input values of $a_0$ and $a_1$.
Moreover, albeit non-negligible, the first-order recoil
correction seems to be moderate for very different input
values of $a_0$ and $a_1$. These two observations lead us to the 
conclusion that rather stringent constraints
on $a_0$ and $a_1$ would emerge, once the $\bar Kd$ scattering length is measured at a reasonable accuracy.

\section{Conclusions}\label{sec:concl}

\begin{itemize}

\item[i)]
In this paper we have studied the first-order recoil correction 
in the multiple-scattering series for the $\bar Kd$ scattering length.
The static $\bar KN$ interactions are treated non-perturbatively and are
summed up to all orders. A perturbative framework is set up for the calculation
of the higher-order recoil corrections (two and more insertions). We plan
to address this issue in our forthcoming publication.

\item[ii)]
The main message which we would like to convey to the reader is that the 
recoil corrections, which, given the large value of the parameter $\xi\simeq 0.5$, 
were {\it a priori} expected to be very large, are in fact 
rather moderate. This conclusion is quite robust and
 holds for all input values
of $a_0,a_1$, randomly taken from the literature.
The smallness
of the recoil corrections can, in part, be attributed to large 
cancellations of the leading contributions in the expansion in $\xi$, which
were discussed above in detail. Partial cancellations could be occurring in the 
sub-leading contributions as well. This issue, however, requires further 
investigation. 

\item[iii)]
The investigation of the first-order recoil corrections is only the
 first step towards the development  of a systematic EFT framework for the 
low-energy $\bar Kd$ system. The next steps would include, in particular, 
estimation of the size of higher-order recoil corrections and a careful
investigation of convergence of the perturbative approach; a study of the role of the 
$\Lambda(1405)$ resonance in the $\bar Kd$ scattering; 
a systematic inclusion of higher-order terms in the effective
Lagrangian such as e.g. the effective range; 
employing the $NN$ potentials
obtained in chiral EFT instead of the phenomenological models and a
systematic inclusion of isospin-breaking effects. 
These studies would provide important steps towards a consistent theory of the low-energy
$\bar Kd$ system which is necessary in order to analyze forthcoming
data on the kaonic deuterium atom~\cite{LNF,JPARK}.

\end{itemize}

\noindent
\textbf{Acknowledgement}
The authors would like to thank A. Cieply, Ulf-G. Mei\ss ner, A. Gal and W. Weise for useful discussions.  We are thankful to Ulf-G. Mei\ss ner for a careful reading of the manuscript.
This work is partly supported by the EU Integrated Infrastructure Initiative 
HadronPhysics3 Project  under Grant Agreement no. 283286. We also acknowledge 
the support by the DFG and NSFC (CRC 110, ``Symmetries and the Emergence of Structure in QCD''). 
One of the authors (AR) acknowledges the support
by the Shota Rustaveli National Science Foundation (Project DI/13/02)
and by Volkswagenstiftung under contract no. 86260.

\renewcommand{\thefigure}{\thesection.\arabic{figure}}
\renewcommand{\thetable}{\thesection.\arabic{table}}
\renewcommand{\theequation}{\thesection.\arabic{equation}}
\appendix

\section{Expansion of the individual amplitudes in powers of $\xi$}
\label{app:exp}

In this appendix, we discuss the expansion of the amplitudes
 ${\cal A}_{0}, {\cal A}_{1}$  and  ${\cal A}^{(c)}$ in powers of $\xi$.
Since the Green function, entering the expression for these amplitudes,
can be written in the following form
\eq
G({\bf p},{\bf l})=\frac{4\pi}{1+\xi/2}\,\frac{1}{{\bf l}^2+2\tilde\xi({\bf p}^2+\gamma^2)}\, ,\quad\quad
\tilde\xi=\frac{\xi}{1+\xi/2}=O(\xi)\, ,
\en
one may argue that the quantity $\tilde\xi$ is an appropriate expansion
parameter. Namely, in the high-energy region we have $l\simeq p$ and the function
$G({\bf p},{\bf l})$ should be Taylor-expanded in the momentum ${\bf p}$. On the opposite, in the low-energy region $l\simeq \tilde\xi^{1/2}p$, the function
$G({\bf p},{\bf l})$ stays intact, whereas other parts of the integrand --
namely, the wave functions and the $NN$ amplitude -- are Taylor-expanded in the momentum ${\bf l}$. It can be easily checked that the convergence of the series
in $\tilde\xi$ is substantially better than in $\xi$, since the latter 
series contains oscillating terms that emerge from expanding 
$\tilde \xi=\xi/(1+\xi/2)$ in powers of $\xi$.

Since the integrals  we  are dealing  with  are  convergent,
  we  are free  to choose  the regularization scheme.  
 Using dimensional regularization,  it can be easily shown that the  expansion  of  ${\cal A}_{0}$  and  $ {\cal A}_{1}$  reads  
  \eq
{\cal A}_{i}=\frac{8\pi}{(1+\xi/2)^2}\,\Big(\sum_{n=1}^{\infty}B_i^n\,\tilde\xi^n
+\sum_{n=0}^{\infty}D_i^n\,\tilde\xi^{\frac{2n+1}{2}}   \Big)\,,
\en
where  $(i=0, 1)$  and
\eq 
B_i^{n}&=&\frac{(-1)^n}{4}\,\int \frac{d^3{\bf p}d^3{\bf l}}{(2\pi)^6}\frac{1}{{\bf l}^2}\Big( \frac{(2({\bf p}^2+\gamma^2))^n}{{\bf l}^{2n}}-\frac{1}{2^n}\Big)\Phi_i({\bf p},{\bf l})\,
\nonumber\\[2mm]
D_i^n&=&\frac{(-1)^{n+1}}{4}\,\int\frac{d^3{\bf p}}{(2\pi)^3}\,\frac{1}{4\pi}\,
(2({\bf p}^2+\gamma^2))^{\frac{2n+1}{2}}\Phi_i^{(2n)}({\bf p})  \,.
\en
Here the  functions  $\Phi_i({\bf p},{\bf l})$ are defined as 
\eq
\Phi_0({\bf p},{\bf l})&=&   \left( \Phi_p({\bf l_1})+\Phi_n({\bf l_1}) \right)\,  
 \left( 
 \Phi_p({\bf l_1})+\Phi_n({\bf l_1})+\Phi_p({\bf l_2})+\Phi_n({\bf l_2})
  \right)\, ,\nonumber \\
\Phi_1({\bf p},{\bf l})&=&   
 \left(\Phi_p({\bf l_1})-\Phi_n({\bf l_1}) \right)\,
 \left(
 \Phi_p({\bf l_1})-\Phi_n({\bf l_1})-\Phi_p({\bf l_2})+\Phi_n({\bf l_2})
 \right)  +2\Phi_x({\bf l_1}) \left( \Phi_x({\bf l_1})-\Phi_x({\bf l_2})\right)\, .
\en
These functions can be  Taylor-expanded as
 \eq\label{phiexp}
  \Phi_i({\bf p},{\bf l})=
\sum_{n=0}^\infty\frac{{\bf l}^{2n}}{(2n+1)2n!}\,
\triangle^n_l\Phi_i({\bf p},{\bf l})\Big|_{{\bf l}=0}\doteq
\sum_{n=0}^\infty {\bf l}^{2n}\Phi^{(2n)}_i({\bf p})\, .
\en
Similarly, the expansion of ${\cal A}^{(c)}$ reads:
  \eq
{\cal A}^{(c)}=\frac{8\pi}{(1+\xi/2)^2}\,\Big(\sum_{n=1}^{\infty}B_c^n\,\tilde\xi^n
+\sum_{n=0}^{\infty}D_c^n\,\tilde\xi^{\frac{2n+1}{2}}   \Big)\,.
\en
In general, explicit expressions for the coefficients $B_c,D_c$ are
very complicated. These, however, simplify somewhat, when the potential $V_{NN}$
 is separable. Below, we give the expressions for the first few coefficients
in case of the Hulth\'{e}n potential, defined in~\ref{app:potentials} 
\eq\label{eq:BDc}
B_c^1&=&\frac{1}{4m_N}\int\frac{d^3{\bf p}d^3{\bf q}d^3{\bf l}}
{(2\pi)^9}\,\frac{\Phi_c({\bf p},{\bf q},{\bf l})}{{\bf l}^4}V_{NN}({\bf p},{\bf q})\,,\nonumber\\[2mm]
B_c^2&=&\frac{1}{4m_N}\int\frac{d^3{\bf p}d^3{\bf q}d^3{\bf l}}{(2\pi)^9}\,
\frac{\Phi_c({\bf p},{\bf q},{\bf l})}{{\bf l}^6}V_{NN}({\bf p},{\bf q})\,
\Big(\frac{\lambda}{16\pi m_N\beta}
-2({\bf p}^2+{\bf q}^2+2\gamma^2) \Big)\,,\nonumber\\[2mm]
D_c^n&=&\frac{1}{4m_N}\int\frac{d^3{\bf p}d^3{\bf q}d^3{\bf l}}{(2\pi)^9}\,
\frac{V_{NN}({\bf p},{\bf q})}{1-A({\bf l})}\frac{{\bf l}^{2n}\Phi_c^{2n}({\bf p},{\bf q})}{({\bf l}^2+2({\bf p}^2+\gamma^2))({\bf l}^2+2({\bf q}^2+\gamma^2))}
\quad\text{~~~for~~~} n=0,1\,.
\en
Here, the  function  $\Phi_c({\bf p},{\bf q},{\bf l})$ is defined as
\eq
\Phi_c({\bf p},{\bf q},{\bf l})=\left( \Phi_p({\bf p+l/2}) + \Phi_n({\bf p+l/2}) \right) \,      \left( \Phi_p({\bf q+l/2}) + \Phi_n ({\bf q+l/2}) \right)\, .
\en
The Taylor expansion of this function at a small ${\bf l}$ takes the form
 \eq\label{phicexp}
  \Phi_c({\bf p},{\bf q},{\bf l})=
\sum_{n=0}^\infty\frac{{\bf l}^{2n}}{(2n+1)2n!}\,
\triangle^n_l\Phi_c({\bf p},{\bf q},{\bf l})\Big|_{{\bf l}=0}\doteq
\sum_{n=0}^\infty {\bf l}^{2n}\Phi^{(2n)}_c({\bf p},{\bf q})\, .
\en
Further, the quantity $A({\bf l})$ in \eqref{eq:BDc} is defined as
\eq
A({\bf l})=\frac{1}{2m_N}\int\frac{d^3{\bf q}}{(2\pi)^3}\,\frac{V_{NN}({\bf q},{\bf q})}{{\bf l}^2+2({\bf q}^2+\gamma^2)}\,.
\en

\section{Nucleon-Nucleon potentials used in the calculations}
\label{app:potentials}

In this section, we specify the nucleon-nucleon potentials used in the calculations. For the kaon, nucleon masses and for the deuteron binding energy the following input values have been used:
\eq
M_K = 493.677\text{~MeV},\quad
m_N=938.92\text{~MeV},\quad
\varepsilon_d = 2.2249\text{~MeV}. 
\en
\subsection{Hulth\'{e}n potential}

The simplest form of the NN potential assumed in this work is the so-called 
Hulth\'{e}n potential. This is a rank-1 separable potential, defined by
\eq
V_{NN}({\bf p},{\bf q})=\lambda \, g({\bf p})g({\bf q})\, ,\quad\quad
g({\bf p})=1/({\bf p}^2+\beta^2)\, .
\en
Here $\beta=1.4 fm^{-1}$ while $\lambda=32\pi m_N\beta\,(\beta+\gamma)^2$ for consistency reasons. In its ansatz 
it is very similar to the more sophisticated phenomenological potentials, such as given in Refs.~\cite{Zankel:1983zz}. While being less realistic than the latter, it contains only long range physics and is therefore 
perfectly suited to study some basic properties of our framework, e.g., the expansion in $\xi$.

\subsection{PEST potential}
 
A more realistic potential is a separable one which is build from the 
Paris potential (see Ref.~\cite{Lacombe:1980dr}) by
 applying the Ernst-Shakin-Thaler method. This potential is therefore referred to as 
the PEST potential in Ref.~\cite{Zankel:1983zz}. It practically  coincides 
with the on- and off-shell behaviour of the Paris potential 
and is given in a fairly simple form:
\begin{align}
V_{NN}({\bf p},{\bf q})=- g({\bf p})g({\bf q})\, ,\quad\quad
g({\bf p})=\sum_{i=1}^6 \frac{C_i}{{\bf p}^2+\beta_i^2}\, ,
\end{align}
where the parameters $\beta_i$ and $C_i$ are given in the table below. It should also be mentioned that the Lippmann-Schwinger equation, which is used in Ref.~\cite{Zankel:1983zz}, has a different normalization than the one
in the present paper:
\eq
(\gamma^2+{\bf p}^2)\Psi({\bf p})=
\frac{m_N}{4\pi}\int d^3{\bf q}V_{NN}({\bf p},{\bf q})\Psi({\bf q})\, .
\en

\setlength\extrarowheight{1pt}
\begin{table}[h]
\begin{center}
\begin{tabular}{|l|c|c|c|c|c|c|}
\hline
			& $i=1$ &$i=2$ & $i=3$ &$i=4$ & $i=5$ &$i=6$\\
\hline
$\beta_i\,[fm^{-1}]$ 	&$1.5$	&$3.0$	&$4.5$	&$6.0$&	$7.5$&	$9.0$\\
\hline
$C_i\,[MeV^{1/2}fm^{-1/2}]$ 	&$3.3786469$ &$-637.41908$&$1750.2432$&$3561.3535$&$-12939.749$ &$8656.6202$\\
\hline
\end{tabular}
\end{center}
\end{table}



\end{document}